\documentclass{aa}  
\usepackage{float}
\DeclareUnicodeCharacter{2212}{-}
\usepackage{graphicx}
\usepackage{txfonts}
\usepackage{hyperref}
\usepackage{xcolor}
\usepackage{appendix}
\usepackage{lscape}
\usepackage{amsmath}
\usepackage{natbib}

\begin{document}

   \title{Search for binary-channel metal-rich RR~Lyrae candidates}

   \author{Hedieh Abdollahi
          \inst{1,2}
          \and
          L\'aszl\'o Moln\'ar\inst{1,3}
          \and
          V\'azsony Varga\inst{1,3,4}
          }

   \institute{Konkoly Observatory, HUN-REN Research Centre for Astronomy and Earth Sciences, MTA Centre of Excellence, Konkoly-Thege Mikl\'os \'ut 15-17, H-1121, Budapest, Hungary\\
             \email{hedieh.abdollahi@csfk.org}           
        \and
   School of Astronomy, Institute for Research in Fundamental Sciences (IPM), Tehran, 19568-36613, Iran
         \and
            ELTE E\"otv\"os Lor\'and University, Institute of Physics and Astronomy, 1117, P\'azm\'any P\'eter s\'et\'any 1/A, Budapest, Hungary
        \and 
            Observatoire de Gen\`eve, Universit\'e de Gen\`eve, Chemin Pegasi 51,
1290 Versoix, Switzerland
             }

\titlerunning{Metal-rich RR~Lyrae candidates in thin-disk}

 
  \abstract
{The existence of dynamically young and metal-rich RR~Lyrae stars challenges conventional notions of  these variable stars. One possible scenario for their formation and evolution is via binary channels involving mass transfer.
This study presents the detection of nine fundamental-mode RR~Lyrae stars residing in the thin disk of the Milky Way with metallicities higher than [Fe/H] > -1.0 dex and showing proper motion anomalies. Our thin disk classification is based on kinematics and supported by $\alpha$-element abundances, where possible. 
We searched for indications of the light-travel time effect (LTTE) in the available literature sources and the TESS photometric data of the stars but found no signs of periodic variations induced by companions within the expected period range. This could be because of a lack of observations as well as sparse measurements and large gaps in the data. We propose a continued search for signs of binarity and a subsequent long-term follow-up of nine targets that satisfy all of our search criteria. Beyond these targets, we also report the detection of slow phase changes in the Blazhko star ST Pic, which could be compatible with the LTTE.}


   \maketitle
%

\section{Introduction}

RR~Lyrae stars are a fascinating group of pulsating stars that have been widely used to study old, metal-poor stellar populations. Traditionally, these stars were thought to belong to ancient stellar systems with ages over 10 Gyr and exhibit a low metal content. Their location in the HR diagram has made them valuable tools for studying stellar pulsations and evolution \citep[see, e.g.,][]{Preston-1964,2015pust.book.....C}. However, recent observations have challenged this traditional view. It seems that some RR~Lyrae stars may not fit the old, metal-poor stereotype. Instead, they exhibit some unexpected characteristics, namley: they are metal-rich and move as young stars do.

Metal-rich RR~Lyrae stars that challenge the canonical scenario and feature [Fe/H] indices reaching or even surpassing the solar value have been identified in the Galaxy \citep[see, e.g.,][]{Layden-1995,2023A&A...674A..18C,2024MNRAS.531..137D}. Further work revealed that these stars are preferentially located in the thin disk and bulge, while the metal-poor and old RR~Lyrae stars are found in the halo and thick disk \citep{Prudil-2020, Zinn-2020,2021MNRAS.502.5686I, 2024MNRAS.52712196B}.
The relationship between the age and metal content distribution of RR~Lyrae variables and the Milky Way disk creates tension with its assumed formation timeline. Traditionally, the thin disk is thought to be younger than the old stellar populations typically associated with RR~Lyrae stars. One way to address this tension is to view metal-rich RR~Lyrae stars as a result of binary evolution.

Metal-poor and old RR~Lyrae stars reach the instability strip through single-star evolution. For low-mass stars, placement on the horizontal branch (HB) is primarily defined by mass-loss and the remaining envelope mass, as well as the opacity profile, both of which are influenced by the star’s metallicity. More metal-rich stars have higher opacities and thus remain cooler and redder, populating the red HB and the red clump \citep{ 2006essp.book.....S, Catelan-2009}. Metal-rich stars would normally require lower initial masses and thus time scales longer than the age of the Universe to evolve into the RR~Lyrae instability strip or, alternatively, higher masses but with significant mass loss to shed enough of their envelopes. Conversely, RR~Lyrae stars originating from interacting binary evolutionary channels are able to cross the instability strip with more metal content and at younger ages. In this scenario, the companion star may remove just enough of the outer envelope of the progenitor during the red giant phase, causing the stripped stars to transition to bluer colors and enter the instability strip, where RR~Lyrae variables are located \citep{2024MNRAS.52712196B}. 

The efficiency of this binary formation channel may be low due to the narrow instability strip, requiring some fine-tuning in the mass transfer for stars to end up there. If the mass transfer is too efficient, stars may end up on the blue HB or even on the extreme HB, crossing the instability strip at a rapid rate. An example of such an object is a low-mass (0.26 M${_\odot}$), RR~Lyrae-like variable in a binary system, OGLE-BLG-RRLYR-02792, which has been classified as a binary evolution pulsator instead \citep{2012Natur.484...75P,2017MNRAS.466.2842K}.

The detection of binary companion candidates can be confidently established through various indicators, including radial velocity (RV) variations, proper motion (PM) anomaly, light-travel time effect (LTTE), or mass transfer \citep[see, e.g.,][]{2016A&A...589A..94L,2016MNRAS.459.4360L, Hajdu-2021}. 
PM anomalies and LTTE can provide candidates for binarity, but confirmation requires additional data such as radial velocities \citep{2022A&A...657A...7K, 2019AJ....157...78J}.
 Mass transfer can significantly alter the evolutionary path and surface composition of the star, making it a potential indicator of binarity. Evidence of mass transfer can be observed through changes in a star's luminosity and spectral features, which are detected via spectroscopic analysis \citep{2014ApJ...783L...8G, 2021AJ....161..248W}. 

Studying the orbital characteristics of binary companions of RR~Lyrae stars is of significant importance. RR~Lyrae stars are known for their pulsations and mass-loss during their evolution, but we know very little about their possible companions. This is because the system has to be distant enough to survive the red giant phase with little to no interaction. Therefore, eclipses would be extremely rare. Furthermore, since RR~Lyrae stars are normally very old, we expect most companions to be either a low-mass K--M dwarf or a white dwarf remnant of a more massive initial companion on a distant orbit, resulting in long and very shallow eclipses. This makes direct detections very difficult. Detections via dynamical effects, such as LTTE or RV variations are also complicated by the presence of the large-amplitude pulsation and the Blazhko effect, the (quasi)periodic modulation of the pulsation \citep{1907AN....175..325B,szabo-2010,smolec-2016}. Photometric surveys have revealed a collection of strong binary candidates \citep{Hajdu-2021}; RV observations, on the other hand, remain inconclusive \citep{Barnes-2021}. 

An investigation of the properties of metal-rich RR~Lyrae stars can provide crucial insights into their origins. Specifically, an inquiry into whether these stars were formed in situ or migrated from other regions requires us to understand their kinematics and chemical abundances. The acquisition and analysis of such data can help answer these fundamental questions.

In this short paper, we aim to investigate whether young and metal-rich RR~Lyrae stars observed in the thin disc exhibit a PM anomaly in their light curves. 
Section \ref{sect:data} provides a detailed overview of the data employed in this study, as well as a comparison of different metallicity estimations relevant to our dataset. Section \ref{sect:results} gives the results in terms of kinematic properties, metallicities and subtypes, $\alpha$-element abundances, and the LTTE. Section \ref{sect:concl} summarizes the results of the paper.

\section{Data}
\label{sect:data}
In this work, we are interested in identifying metal-rich, thin-disk RR~Lyrae stars that could have formed through binary mass transfer, following the proposal of \citet{2024MNRAS.52712196B}.


\cite{2021MNRAS.502.5686I} applied selection criteria to the RR~Lyrae populations based on a multi-component kinematic model. To infer the behavior of the velocity ellipsoid between approximately 3 and 30 kpc from the center of the Galaxy, they used PM measured by \textit{Gaia} and imposed a four-fold symmetry, which found two populations: a non-rotating halo-like population, which could be described as a superposition of isotropic and radially-biased parts, and a rotating disc-like population, which was much smaller and had a mean azimuthal speed and velocity dispersion consistent with a young and metal-rich thin disc stellar population. We started our selection with this sample of 2,143 stars. 

Given the expectation that these stars evolve within binary systems, it is crucial to examine the influence of their binary companions. One potential direct effect under investigation is on the PM of these RR~Lyrae stars. 
PM anomalies between HIPPARCOS and \textit{Gaia} measurements are discussed in  \cite{2019A&A...623A.116K}, whereas \cite{2019A&A...623A.117K} looked for spatially resolved common PM pairs, indicating wider and potentially gravitationally bound systems. The two studies analyzed nearly a thousand targets combined. We refer to all of them as having a PM anomaly for the sake of brevity from here onwards.
We cross-matched both the strong detections and the candidates with the disk population of \citep{2021MNRAS.502.5686I}. Out of the 2,143 stars, we identified only 24 RR~Lyrae stars that display a PM anomaly.

\begin{figure*}{}
    \centering
   {\includegraphics[width=\textwidth]{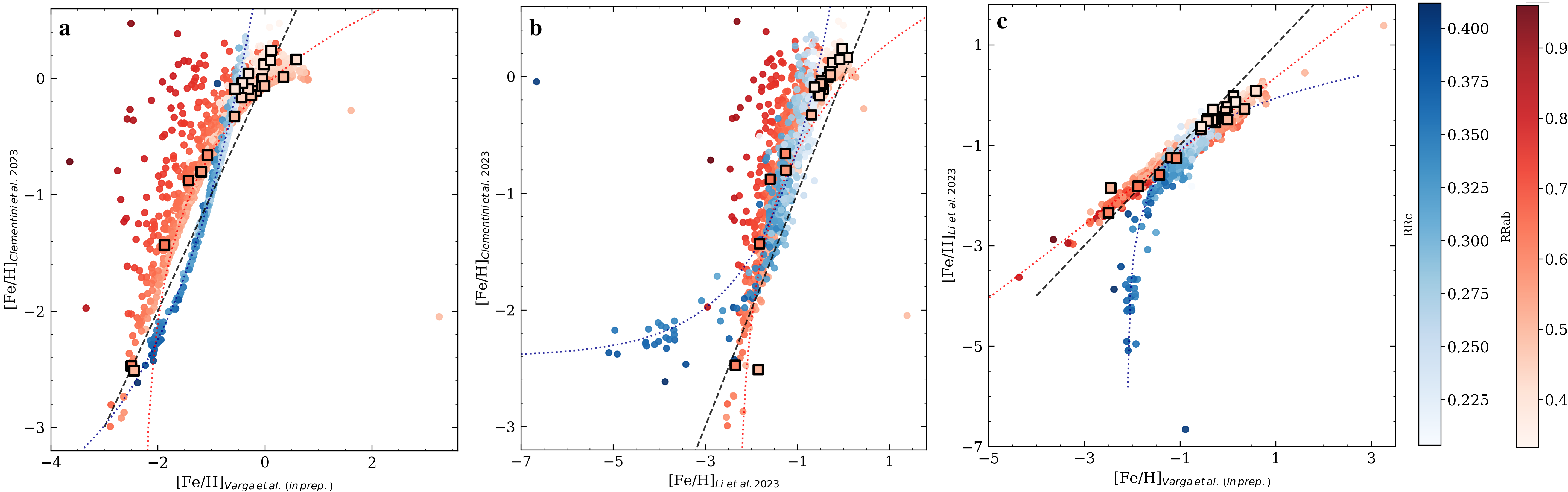}} 
 
    \caption{Comparison of metallicity distribution for RR~Lyrae stars with different calibrations of \cite{2023ApJ...944...88L}, \cite{2023A&A...674A..18C}, and Varga et al. (in prep.). The distributions of disk stars (circles) and stars with PM anomalies (squares) are categorized by color for RRab and RRc types. The color bars are based on the periods derived from \textit{Gaia} DR3 \citep{2023A&A...674A..18C}.}
    \label{fig:Metallicity}
\end{figure*}

\subsection{Photometric metallicity}
\label{subsec:metallicity}
Since chemical composition affects the envelope structures of stars, it also affects stellar pulsation modes propagating within the envelope. Metal-rich RR~Lyrae stars, for example, generally have shorter periods. The metallicities of RR~Lyrae stars thus can be estimated from the light curves via photometric methods. Light curve shapes are quantified via relative Fourier-parameters, defined as $R_{i1} = A_i/A_1$ and $\phi_{i1} = \phi_i - i\,\phi_1$ by \citet{simon-1982}. However, these methods rely strongly on calibrations based on their spectroscopic metallicities. 

To further study the differences in photometric metallicities, we developed a new calibration method specifically tailored for \textit{Gaia} \textit{G} photometric data, encompassing both the RRab and RRc subtypes (Varga et al. (in prep.)). 
A detailed description of our calibration methodology can be found in Appendix \ref{sec:Our own metallicity calibration}.

Figure \ref{fig:Metallicity} presents a comparison of three different photometric metallicity calibrations for RR~Lyrae stars. The disk stars that are in common between the comparative surveys in subplots a, b, and c are represented by the circles. These data contain 1,379 RRab and 339 RRc stars from the disk sample for which $\phi_{31}$ are available in \textit{Gaia}. The squares in these plots highlight RR~Lyrae stars with PM anomalies within this sample. The stars are color-coded to indicate their \textit{Gaia} DR3 periods \citep{gaia-dr3,2023A&A...674A..18C}. In the panels, the blue and red dotted lines represent polynomial fits for the RRc and RRab stars, respectively, alongside dashed black identity lines. It is evident that the RRab populations in each plot align closely with the identity lines, suggesting a good agreement. In contrast, deviations from the identity lines for RRc stars point to systematic discrepancies between the catalogs at low metallicities. 
In subplots a and b of Figure \ref{fig:Metallicity}, \textit{Gaia} DR3 consistently overestimated metallicities compared to the \cite{2023ApJ...944...88L} and Varga et al.\ (in prep.) values. Nevertheless, the least deviation from identity, considering both RRab and RRc results, occurs between the \cite{2023A&A...674A..18C} and Varga et al.\ (in prep.) calibrations.

The RRc outliers in Figure \ref{fig:Metallicity} subplots b and c need further investigation. These are the shortest-period and thus hottest RRc stars. The low metallicities and the high $T_{\rm eff}$ means that iron lines will be very weak in these stars, which makes accurate spectroscopic [Fe/H] determinations challenging. Furthermore, the lowest RRc metallicity values are based on extrapolations of the relations, which can become subject to uncertainties rather easily. Recently, \citet{Muraveva-2025} also found a similar discrepancy with the \cite{2023ApJ...944...88L} RRc metallicities. However, we are interested in metal-rich stars in this work, so this discrepancy in the RRc calibrations does not affect our results directly.

   \begin{figure}
   \centering   \includegraphics[width=1\columnwidth]{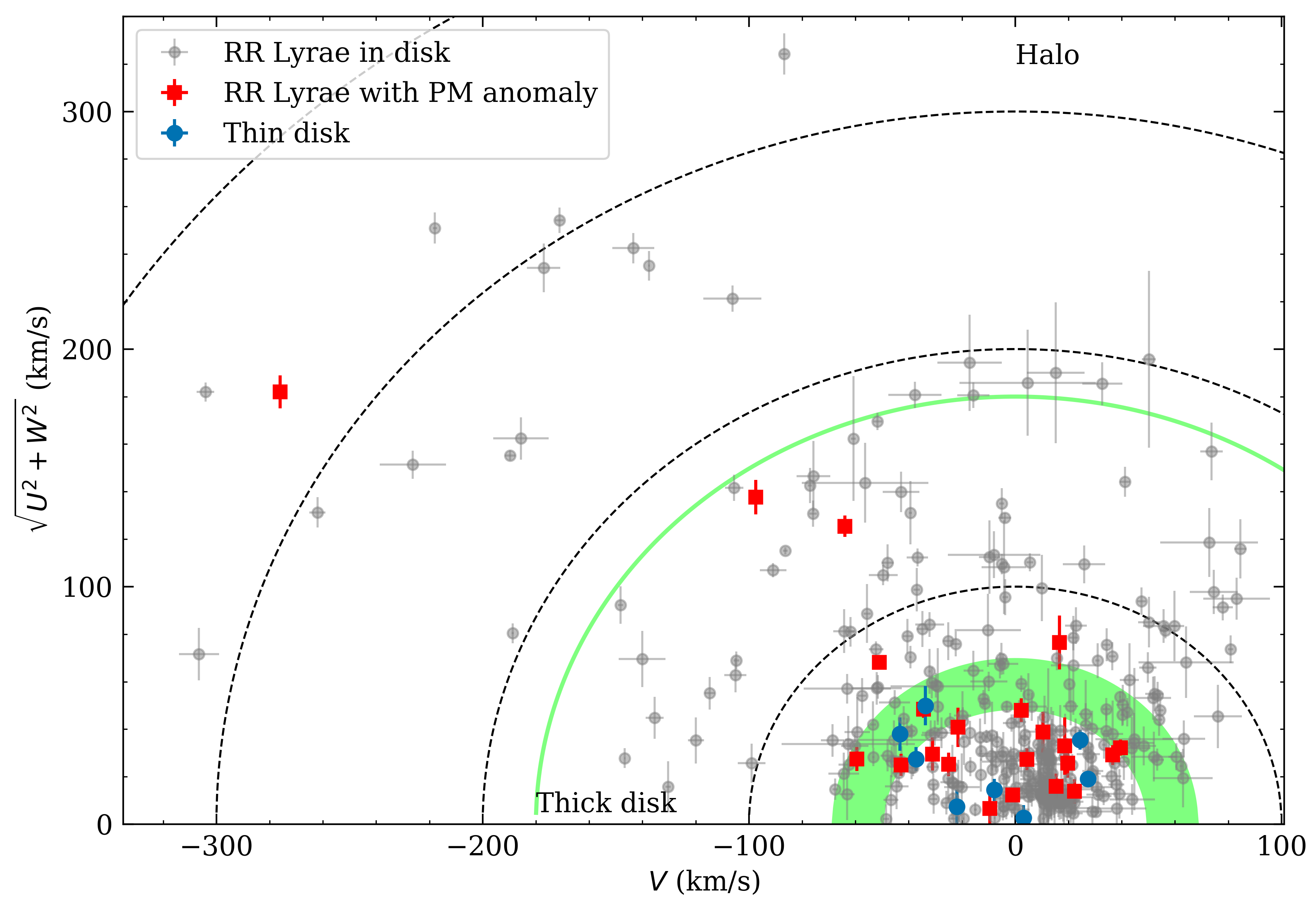}
      \caption{Toomre diagram shows gray circles for RR~Lyrae stars from the \cite{2021MNRAS.502.5686I} population and red squares for those with PM anomalies found by \cite{2019A&A...623A.117K, 2019A&A...623A.116K}. The blue circles represent a population of the thin disk classified by their $\alpha$-element abundances \citep{2023ApJ...958...32D, 2024MNRAS.52711082D}.}
         \label{fig:Toomre}
   \end{figure}

\begin{figure}{}
    \centering
   {\includegraphics[width=\columnwidth]{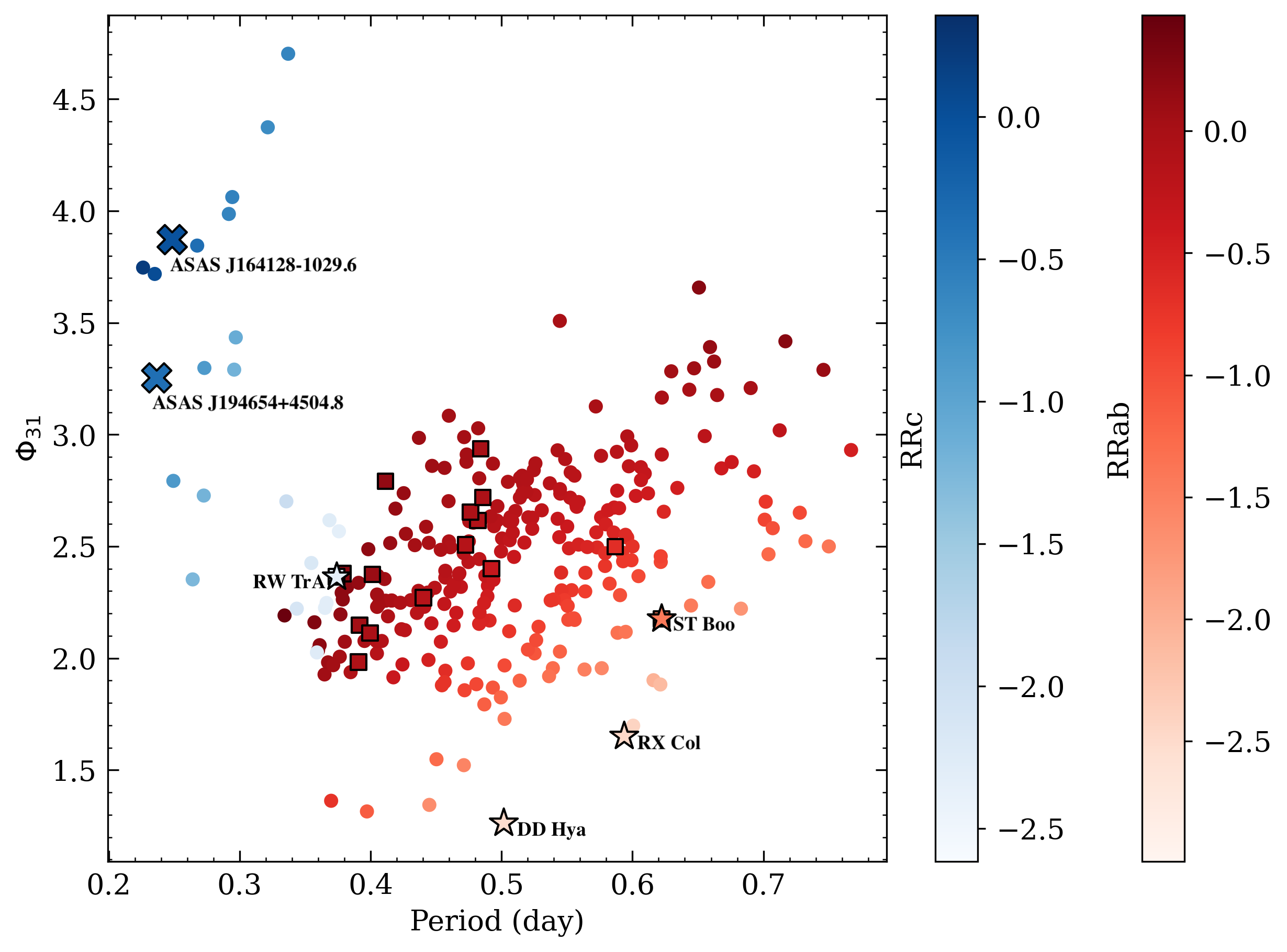}} 
    \caption{ $\phi_{31}$ Fourier phase parameter \textit{vs.} pulsation period is shown for RR~Lyrae stars in the thin disk (circles). Data includes stars with PM anomalies in the thin disk (squares) and the most metal-poor stars (stars). 
    Blue crosses are highlighted the most metal-rich ([Fe/H] > -0.5 dex) RRc in the thin disk. The color bars represent metallicity data from  \textit{Gaia} DR3 \citep{2023A&A...674A..18C}.}
   \label{fig:phi-Period}
\end{figure}

\section{Results}
\label{sect:results}
We employed three criteria to discern RR~Lyrae stars that could have originated from binary channels: metallicity, as they tend to be metal-rich; $\alpha$-elements, as the thin disk is $\alpha$-poor relative to their iron content; and velocities, which is also an indicator for thin disk populations. 
Of the 24 RR~Lyrae stars, two do not have metallicity data (CG Peg and DH Peg) and four are classified as metal-poor with [Fe/H] < -1 dex. UY Cyg and SW And do not have available RV information.

\subsection{Kinematics}
In Figure \ref{fig:Toomre}, the Toomre diagram shows the calculated velocities in the Cartesian X and Y directions.  
Disk RR~Lyrae stars identified by \citet{2021MNRAS.502.5686I} are marked with gray circles, while stars that also show PM anomalies are the red squares. The blue circles represent a population of thin disks classified by their $\alpha$-element abundances. We expect to observe young and metal-rich stars within the thin disk, characterized by a total velocity of less than 50 km s$^{-1}$. Stars with total velocities ranging from 50 to 70 km s$^{-1}$ occupy a zone where the populations of the thin and thick disks overlap \citep{2024MNRAS.52711082D, 2023ApJ...958...32D}.
The heliocentric coordinates, in this case, are located at V$_\odot$ = (12.9, 245.6, 7.78) km s$^{-1}$, following the method of \cite{2022ApJS..258....8M}. The PM and RV values are from \textit{Gaia} DR3.

\subsection{Metallicities and subtypes}
\label{subsec:Metallicities and subtypes}
Figure \ref{fig:phi-Period} depicts the distribution of $\phi_{31}$, the phase difference between the pulsation frequency and the second harmonic of the light-curve, relative to the pulsation period for the thin-disk population. The circles in Figure \ref{fig:phi-Period} include 286 disk population stars from \cite{2021MNRAS.502.5686I}, specifically selecting those with an available $\phi_{31}$ parameter in \textit{Gaia} DR3 and a total velocity of less than 70 km s$^{-1}$. Within this dataset, there are 262 RRab and 24 RRc stars.
Thin-disk stars with PM anomalies are indicated with squares, while the four most metal-poor stars in our sample (RW TrA, ST Boo, DD Hya, and RX Col) as presented in Table \ref{table:RR~Lyrae} are marked with star symbols. Considering the velocity criteria and the availability of the $\phi_{31}$ parameter, we are left with 16 RR~Lyrae stars with PM anomalies.
The color bars differentiate between RRab and RRc variables, based on their metallicities, as derived from \cite{2023A&A...674A..18C}. 

RW TrA is located within the thin disk based on velocity criteria (V < 70 km s\(^{-1}\)) but it appears to be very metal-poor if we accept the RRc classification in \textit{Gaia} DR3. However, it is near the region of RRab variable stars in Figure \ref{fig:phi-Period}, suggesting a potential misclassification, which can happen to the shortest-period RRab stars \citep{2016IBVS.6175....1M}. We reanalyzed the \textit{Gaia} DR3 \textit{G}-band photometry of the star and found the $\phi_{31}$ value to be about one $\pi$ off from the DR3 value, at $\phi_{31}^s = 5.514 \pm 0.085$. If we recalculate the metallicities with this value, we arrive at [Fe/H] = +0.15, and RW~TrA becomes a metal-rich, short-period RRab star, fitting a metal-rich thin disk profile. This agrees not only with earlier RRab classifications, but with its spectroscopic metallicity as well ([Fe/H] = +0.12, \citet{Crestani2021}).
Thus, all classifications in Table \ref{table:RR~Lyrae} are from \citet{2023A&A...674A..18C}, except for RW TrA, for which we use the earlier RRab classification.

In Figure \ref{fig:phi-Period}, two RRc variables, denoted by blue crosses, are highlighted based on their high metallicity ([Fe/H] > -0.5 dex) and brightness exceeding 13 mag in the \textit{G} band. The characteristic positions of these stars as outstanding candidates for follow-up observations with  Transiting Exoplanet Survey Satellite (TESS). These stars may exhibit PM anomalies in their astrometry. The PM anomalies could potentially reveal the existence of a component for these stars. Moreover, \citet{2024MNRAS.52712196B} predicts that binary interactions produce low-mass RR~Lyrae stars; in addition, for RRc stars, we now have the possibility to estimate seismic masses, following the technique developed by \citet{Netzel-2022,Netzel-2023}. Table \ref{table:RRc~Lyrae} details their Name, Source ID and metallicity from \cite{2023ApJ...944...88L, 2023A&A...674A..18C} and Varga et al. (in preparation).

   \begin{figure}
   \centering
   \includegraphics[width=1\columnwidth]{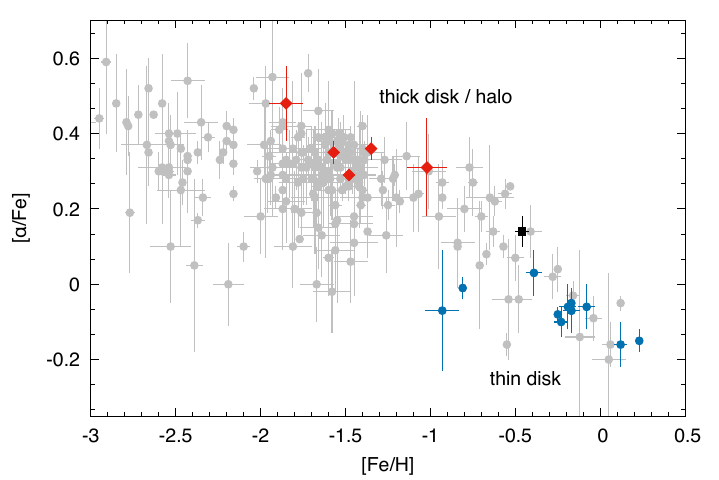}
      \caption{[$\alpha$/Fe] versus [Fe/H] abundances of bright RR~Lyrae stars, based on \citet{Crestani2021} (gray circles). The blue and red points correspond to the thin and thick stars mutual between the works of \citet{Crestani2021} and \citet{2021MNRAS.502.5686I}, present in our study. The star marked with the black square has an unclear disk classification.}
         \label{fig:Alpha}
   \end{figure}

\subsection{$\alpha$-element abundances}
\label{subsec:alpha-element}
Since the thin and thick disks overlap kinematically, we need to include another constraint to fully separate the two populations. A strong indicator is the chemical composition, namely, the abundances of $\alpha$-elements relative to the [Fe/H] indices, as the thick disk is $\alpha$-rich, while the thin disk is $\alpha$-poor relative to their iron content \citep[see, e.g.,][]{2003A&A...410..527B, 2005A&A...438..139S, Imig_2023}. A homogeneous spectroscopic survey was recently undertaken by \citet{Crestani2021} who determined the $\alpha$-element abundances for 162 RR~Lyrae stars. We cross-matched this with the disk population sample and found 18 matches.

In our sample of 24 stars with a PM anomaly, there are nine from the thin disk and two from the thick disk. Figure \ref{fig:Alpha} shows the correlation between [$\alpha$/Fe] and [Fe/H], indicating a clear distinction between the two groups. The gray circles represent the population in \citet{Crestani2021}, while the blue and red points highlight the mutual thin and thick stars in our sample, respectively. The black square falls close to the divisions present in the literature, making its classification unclear \citep[see, e.g.,][]{Adibekyan-2012,Recio-2014,Vincenzo-2021}. Eleven members of this sample are experiencing PM anomalies.

\begin{figure}{}
    \centering
   {\includegraphics[width=0.5\textwidth]{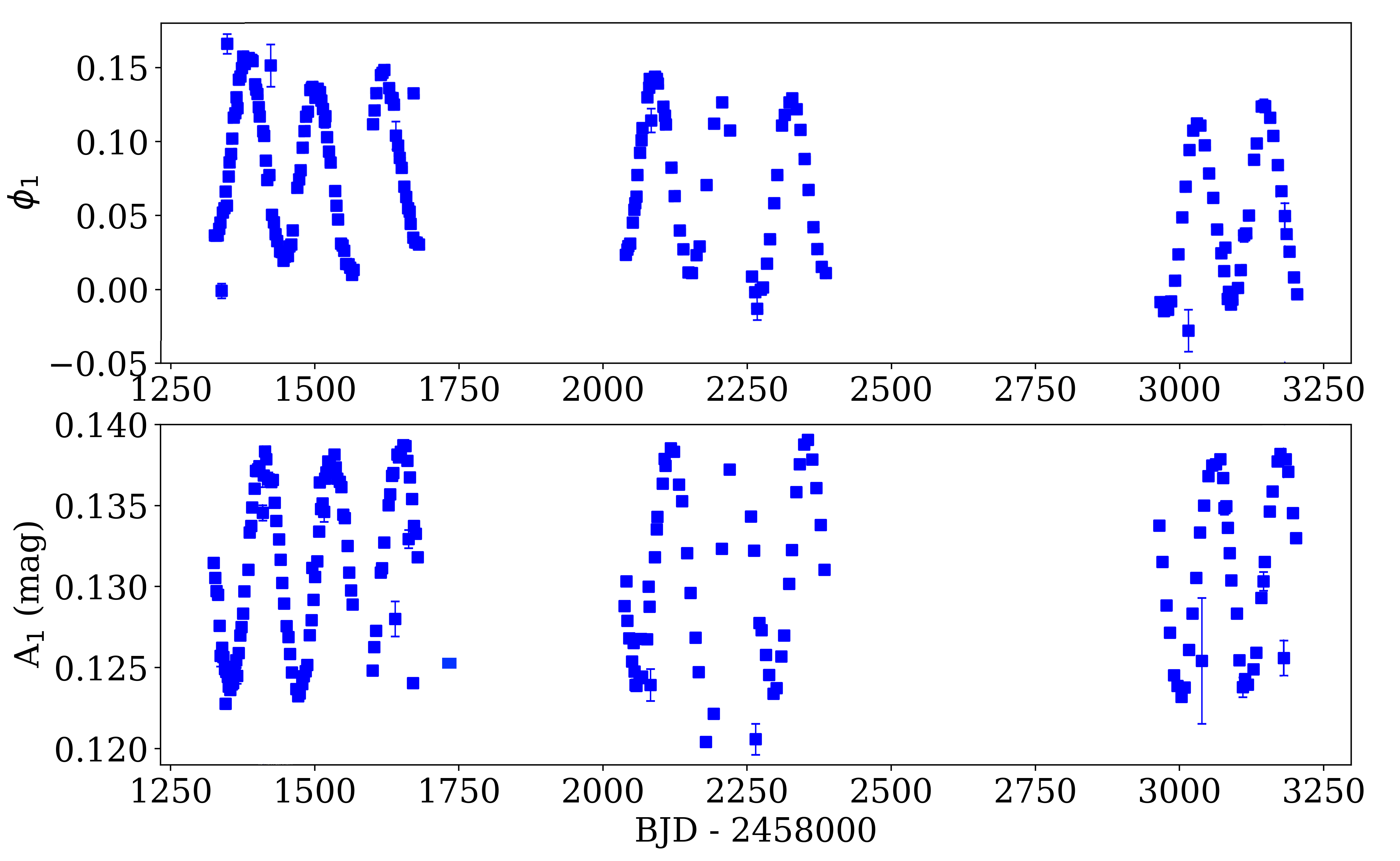}} 

    \caption{Temporal variation in phase ($\phi_{1}$) and amplitude ($A_{1}$) of the pulsation frequency of ST Pic from the TESS survey. A long-term variation is present in the phase which is not seen in the amplitudes.}
    \label{fig:LC}
\end{figure} 

\subsection{Light-travel time effect}

One clear indication of binarity for our candidates would be the detection of the LTTE in the pulsation. \citet{2024MNRAS.52712196B} predict orbital periods to be in the 1--2000\,d range for systems that can form a metal-rich RR~Lyrae star. Since such orbits are wide, we expect the O-C signals to be detectable, assuming a long enough temporal coverage. Subtle signs of a companion star emerge only with long-term observations \citep{2016MNRAS.459.4360L, 2024A&A...691A.108S}.

We collected photometric data for our 24 targets from the TESS mission \citep{2015JATIS...1a4003R}. A simple Fourier series with three frequency components is sufficient for modeling RRc stars, which typically have less complexity in their light curves, which helps to avoid overfitting. We utilized a higher order Fourier series with 8-12 frequency components to account for the more complex light-curves of RRab stars.
We calculated the Fourier parameters, including amplitudes and phases, for these data sets, cutting them into segments of various lengths depending on the number of data points in each data set. 

To ensure an accurate analysis of the light curves, we cross-matched the RR~Lyrae star catalog from \cite{2021MNRAS.502.5686I} with candidate binary stars from the RRLyrBinCan database\footnote{\url{https://rrlyrbincan.physics.muni.cz/}} \citep{2016MNRAS.459.4360L, 2016CoKon.105..209L}. Thus, we identified two stars with independent indications: AQ Lyr and CN Lyr.  

AQ Lyr shows cyclic O-C variations but does not belong to the PM anomaly sample and the estimated companion mass appears to be too high \citep{2016MNRAS.459.4360L}. CN Lyr resides in the thin disk based on $\alpha$-elements classification and showed signs of changes in its systemic RV over the years, but \citet{Barnes-2021} concluded that these are not significant.

Furthermore, RR~Lyrae stars with the Blazhko effect show irregular light-curve modulation over long periods.
The stars ST~Pic, Z Mic, ST Boo, SW And, DD Hya, and CZ Lac, identified from the total of 24 RR~Lyrae stars, demonstrate the Blazhko effect in their pulsation period. The observed Blazhko periods for these stars are between 284 and 14.6 days, which is considerably shorter than the expected orbital periods \citep{2013A&A...549A.101S}.

Figure \ref{fig:LC} illustrates the variations in phase ($\phi_1$) and amplitude (A$_1$) over time for one of the Blazkho stars, ST Pic. The TESS observations were collected with different integration times during the observation cycles; here we show the analysis of the 120~s cadence data. Both the amplitude and phase variations clearly show the Blazhko modulation. However, the phase variations suggest the presence of a longer cycle as well, which is not present in the amplitude data. This raises the possibility of a LTTE on top of the Blazhko effect. However, period changes and other factors such as mode-switching, internal structure changes, or magnetic activity may also influence phase and amplitude variations and could be more commonly observed than LTTE. Period changes in RR~Lyrae stars may occur from evolutionary changes and interactions with companion stars. These variations can be observed as gradual or abrupt shifts in the pulsation period \citep{2016MNRAS.461.1032A, 2018ApJ...863..151L,Csornyei-2022}. 
Furthermore, random fluctuations in cycle lengths can lead to quasi-periodic or irregular O−-C value variations, even without actual period changes. The C2C variation is particularly significant for RR~Lyrae stars, possibly due to turbulent convection. New statistical models and Fourier analysis indicate that C2C variations could explain O−C curves without mean period changes (Benk\H{o} et al., in prep.).

We note that despite the higher temporal resolution, the precision of the phase variations in our study and the study of \citet{2022ApJS..258....8M} appear to be sufficient to detect the O-C variations reported by \citet{Hajdu-2021}, suggesting that similar signals would be recoverable from the TESS observations.

\section{Conclusions}
\label{sect:concl}
In this work, we searched for viable candidates for the mechanism proposed by \citet{2024MNRAS.52712196B} to explain the existence of metal-rich, dynamically young RR~Lyrae stars. We identified 24 stars that show PM anomalies or are members of co-moving pairs. Among this population, 18 are classified as metal-rich ([Fe/H] < -1 dex) and 13 of these are identified within the thin disk based on the velocity criteria (V < 70 km s$^{-1}$, \citep{2023ApJ...958...32D, 2024MNRAS.52711082D}). Furthermore, the $\alpha$-element abundances are considered \citep{Crestani2021}. According to this criterion, among the 24 RR~Lyrae exhibiting PM anomalies, 9 are located within the thin disk, while 2 are situated in the thick disk. The thin disk contains nine stars that satisfy all three criteria: AA~CMi, CN~Lyr, DX~Del, FW~Lup, HH~Pup, RW~TrA, SW~And, TW~Her, and U~Pic. The evolution of these stars could have been governed by binary evolutionary channels.

However, we did not observe any significant variation in the available photometric data of the targets in the predicted 1--2000 d period range that would indicate the presence of a binary system. The only candidate that matches two of our criteria and where we see signs of a possible slow O--C signal is ST~Pic, but it was not observed by \citet{Crestani2021}. We also highlight two RRc stars that fit some of our criteria and could be targets of asteroseismic analysis through their extra modes.

Further TESS observations could alleviate the lack of definite O--C detections, but a long-term RV follow-up would also be required to unambiguously identify metal-rich RR~Lyrae stars as products of binary interactions. While some RR~Lyrae stars have been followed up on with some degree of regularity, existing long-term RV data are still very sparse \cite{Barnes-2021}. We propose setting a focus  on the targets given in Tables \ref{table:RR~Lyrae} and \ref{table:RRc~Lyrae} in our Letter as the best candidates for future studies.

\begin{acknowledgements}
We acknowledge with thanks the comments of the anonymous referee that helped us to improve the paper. This research was supported by the `SeismoLab' KKP-137523 \'Elvonal grant and the NKFIH excellence grant TKP2021-NKTA-64 of the Hungarian Research, Development and Innovation Office (NKFIH). H.A.\ thanks the hospitality of Konkoly Observatory and the Research Centre for Astronomy and Earth Sciences during her visit to Budapest, where part of this project was carried out. V.V.\ has been supported by the undergraduate research assistant program of Konkoly Observatory. This research has made use of NASA’s Astrophysics Data System.
\end{acknowledgements}

%
\bibliographystyle{aa} 

\bibliography{Arxiv}  
\appendix

\section{Photometric metallicity}
\label{sec:metallicity compariosn}
\subsection{Comparison of the calibrations}
The comparison of metallicity estimation in \cite{2023ApJ...944...88L}, \cite{2023A&A...674A..18C}, and Varga et al. (in prep.) are presented in Figure \ref{fig:Metallicity}. All these methods have utilized data from \textit{Gaia} DR3, but they employed different techniques and calibrations. \cite{2023ApJ...944...88L} utilized calibrated absolute magnitude-metallicity relations and NIR period-absolute magnitude-metallicity relations, while the second study employs the period and Fourier parameters of the G-band light-curves for this estimation. These differences may be due to the various parameters used for the calculation. \cite{2023ApJ...944...88L} used both the amplitude ratio and the phase difference, whereas \cite{2023A&A...674A..18C} mainly relied on the phase difference to estimate metallicity. However, these differences are largely caused by the calibration processes themselves.
The \textit{Gaia} DR3 are not to be relied upon for RRab periods longer than 0.7 days, as the relation established by \citet{Nemec-2013} and used by \cite{2023A&A...674A..18C} has not been calibrated for stars beyond that; also, the extrapolation of the second-order terms lead to artificially high metallicities there.

\subsection{Our own metallicity calibration}
\label{sec:Our own metallicity calibration}
To investigate the differences in photometric metallicities further, we calculated them with our calibration as well. While the details of the work are to be published elsewhere (Varga et al., in prep.), the key aspects are presented below. 

We established a novel photometric metallicity relation in the \textit{Gaia G} photometric band for both the RRab and the RRc subtypes. The calibration data consisted of the photometric parameters of the \textit{Gaia} DR3 RR~Lyrae sample \citep{2023A&A...674A..18C}, and spectroscopic $\mathrm{[Fe/H]}$ measurements from various sources for both subtypes: the Apache Point Observatory Galactic Evolution Experiment (APOGEE) Data Release 17 (RRab stars only, \citealt{apogee-sdss-dr17}), \citet{C21} (RRab stars only), the GALactic Archaeology with HERMES survey (GALAH) Data Release 3 (RRab and RRc stars, \citealt{galah-dr3}), and the Carnegie RR~Lyrae Survey (CARRS; RRc stars only, \citealt{Sneden-2018-carrs}). Spectroscopic $\mathrm{[Fe/H]}$ values in all calibration data sets were brought to the solar reference $\log \epsilon(\mathrm{Fe}) = 7.50$ of \citet{asplund-2009-solar-abundances}, also used by \citet{2023A&A...674A..18C}.

Estimating the $\mathrm{[Fe/H]}$ values using only photometry is approached in an empirical way, based on the period of the dominant pulsation mode, $P$ (fundamental mode for RRab, first overtone for RRc stars) and the shape of the light-curve in \textit{Gaia G} band described by its Fourier phase difference ($\varphi_{31}\equiv \varphi_3-3\varphi_1$). For the form of the fitting model $\mathrm{[Fe/H]} = f(P, \varphi_{31}) $, we chose $\mathrm{[Fe/H]}_\text{RRab} = a + b\cdot P + c\cdot \varphi_{31} + d\cdot P^2 $ and $\mathrm{[Fe/H]}_\text{RRc} = a + b\cdot P + c\cdot \varphi_{31} $, respectively, for the RRab and the RRc subtypes, after investigating the importance of each term in a second-degree polynomial model. However, in order to avoid biases, especially in the uncertainty estimation, models were not directly fitted, but each metallicity value was individually calculated by randomly varying the input photometric parameters and the calibration data sets according to each uncertainty level, also accounting for the intrinsic deviation of the empirical method itself.

The coverage of the calibration data sets on the $P$--$\varphi_{31}$ parameter space is satisfactory, but significant systematic deviations can be expected outside the range of spectroscopic metallicity values covered: $\mathrm{[Fe/H]}_\text{RRab, calib.}\in [-3.1, 0.1]$, $\mathrm{[Fe/H]}_\text{RRc, calib.}\in [-2.0, -0.2]$. 
In \citet{Netzel-2024}, a similar relation for estimating photometric metallicities for RRc stars in the TESS passband is introduced and briefly described. This relation, which will also be incorporated into Varga et al.\ (in prep.), extends to low RRc metallicities. In particular, within the low metallicity range, below approximately -1.5 dex, it yields [Fe/H] values that reside between our Gaia RRc measurements and those calculated by \citet{2023ApJ...944...88L}. This indicates that the true metallicity values may likely be situated within this interval. While the metallicities reported by \cite{2023A&A...674A..18C} and low-metallicity RRc findings by Varga et al. (in prep.) align with those of \cite{2023ApJ...944...88L}, it is possible that the first two assessments are inaccurate and the findings of \cite{2023ApJ...944...88L}  may also be subject to error.

\section{Data tables}
Table \ref{table:RR~Lyrae} lists 24 RR~Lyrae stars with PM anomalies. It includes the star's Name, Source ID, G magnitude, type \citep{2023A&A...674A..18C}, and photometric metallicities with errors extracted from \cite{2023ApJ...944...88L, 2023A&A...674A..18C} and Varga et al. (in prep.). The separation of stars in three sections in Table \ref{table:RR~Lyrae} is based on the $\alpha$-element abundances, which are discussed in Subsection \ref{subsec:alpha-element}. A detailed explanation of metallicity estimation by Varga et al. (in prep.) is presented in Appendix \ref{sec:Our own metallicity calibration}, along with a comparison of different metallicity calibrations in Subsection \ref{subsec:metallicity} and Appendix \ref{sec:metallicity compariosn}.
The characteristics related to RRc are presented in Table \ref{table:RRc~Lyrae}, highlighting the foremost metal-rich stars in our sample.

\begin{table*}
\caption[]{Metallicity comparison utilizing different calibrations for RR~Lyrae stars with a PM anomaly.}
\label{table:RR~Lyrae}     
\begin{tabular}{lllllll}
\hline\hline

Name & Source\_id & G (mag) & [Fe/H]$_{Varga\,in\, prep.}$ &   [Fe/H]$_{Li\,et\,al.\,2023}$ &   [Fe/H]$_{DR3}$ &Type\\

 \hline
\multicolumn{7}{l}{\textbf{Population in the thin disk}} \rule{0pt}{10pt}\\ \hline
AA CMi & 3111925220109675136 & 11.618&-0.055$\pm$0.394&-0.460$\pm$0.330&-0.067$\pm$0.242&RRab\\
CN Lyr & 4539434124372063744 & 11.297&0.588$\pm$0.330& 0.080$\pm$0.280&0.163$\pm$0.218&RRab\\
DX Del & 1760981190300823808&9.914&-0.259$\pm$0.270&-0.550$\pm$0.250&-0.144$\pm$0.218&RRab\\
FW Lup $^{\star}$& 6005656897473385600 & 8.934 & 0.346$\pm$0.349&-0.280$\pm$0.290&0.013$\pm$0.219&RRab\\
HH Pup & 5510293236607430656 &11.344& -0.556$\pm$0.355&-0.630$\pm$0.300&-0.090$\pm$0.265&RRab\\
RW TrA $^{\star}$ \ & 5815008831122635520 & 11.287 & 0.158$\pm$0.305 &-0.140$\pm$0.214&0.151$\pm$0.026 & RRab\\
SW And & 2857456211775108480&9.705&-0.046$\pm$0.310&-0.340$\pm$0.280&-0.006$\pm$0.226&RRab\\
TW Her & 4596935593202765184 & 11.393&-0.420$\pm$0.366&-0.480$\pm$0.310&-0.038$\pm$0.252&RRab\\
U Pic & 4784552718312266624 & 11.548 & -0.436$\pm$0.354&-0.520$\pm$0.300&-0.165$\pm$0.246&RRab\\
\hline\hline
\multicolumn{7}{l}{\textbf{Population outside the thin disk}} \rule{0pt}{10pt} \\ \hline
DH Peg & 2720896455287475584 & 9.517 & - &  - &- & -\\
ST Boo&1374971558625266432&11.073&-1.876$\pm$0.510&-1.820$\pm$0.390&-1.433$\pm$0.764&RRab\\
Z Mic & 6787617919184986496 & 11.524&-1.073$\pm$0.332&-1.260$\pm$0.290&-0.660$\pm$0.271&RRab\\

\hline\hline
\multicolumn{7}{l}{\textbf{Unspecified alpha element criteria }} \rule{0pt}{10pt} \\ \hline
CG Peg & 1797739517580809856 & 11.147 & -&  - & -& - \\
CZ Lac & 2000976545403561984&11.629&-0.314$\pm$0.367&-0.500$\pm$0.310&-0.092$\pm$0.248&RRab\\
DD Hya & 3090871397797047296&12.344&-2.445$\pm$1.384& -1.850$\pm$0.990&-2.513$\pm$2.396&RRab\\
EW Cam &1111846056593351168&9.396&-1.429$\pm$0.273&-1.590$\pm$0.260&-0.879$\pm$0.234&RRab\\
EZ Cep $^{\star\star}$ &568205142263138304 & 12.754&0.117$\pm$0.402&-0.030$\pm$0.320&0.239$\pm$0.225&RRab\\
RX Col & 2884776807984012160&12.545&-2.494$\pm$0.470&-2.350$\pm$0.380&-2.474$\pm$0.651&RRab\\
ST Pic $^{\star}$& 5481339590753150208 & 9.452 & -0.007$\pm$0.315&-0.490$\pm$0.280&-0.064$\pm$0.225&RRab\\
TW Lyn&924355886418231680&11.831&-0.162$\pm$0.571&-0.440$\pm$0.440&-0.108$\pm$0.300&RRab\\
TZ Aur&949205983077666944&12.010&-0.310$\pm$0.809&-0.290$\pm$0.570&0.043$\pm$0.441&RRab\\
UY Cyg & 1858568795812429056&11.018&-1.185$\pm$0.295&-1.250$\pm$0.270
& -0.802$\pm$0.248&RRab\\
V0690 Sco & 4035521829393903744 & 11.493 & -0.567$\pm$0.314 & -0.329$\pm$0.280 & -0.690$\pm$0.250 & RRab\\
V0830 Cyg & 1873027231966783232 &13.234&-0.015$\pm$0.414&-0.240$\pm$0.340&0.120$\pm$0.235&RRab\\
\hline
\end{tabular}
\tablefoot{The stars in the first section belong to the thin disk according to the $\alpha$-elements criteria. The classifications of RR~Lyrae stars are sourced from \cite{2023A&A...674A..18C}. CG Peg and DH Peg are not classified as variables and lack LC data in the \textit{Gaia} RR~Lyrae catalog, so their types are not reported. They are classified as  RRab and RRc in \cite{2019A&A...623A.117K, 2019A&A...623A.116K}, respectively.
\\ $^{\star}$ Do not have specified types in \cite{2019A&A...623A.117K,2019A&A...623A.116K}. Refer to \ref{subsec:Metallicities and subtypes} for more information on RW TrA. \\ 
$^{\star\star}$ Different types reported in  \cite{{2019A&A...623A.117K,2019A&A...623A.116K}}.
}

\end{table*}

\begin{table*}
\caption[]{
\label{table:RRc~Lyrae} Further potential thin disk RRc-type stars in the sample available for seismic analysis.}
\begin{tabular}{llllll}
     \hline\hline

Name &Source\_id & G (mag) & [Fe/H]$_{Varga\,in\, prep.}$ &  [Fe/H]$_{Li\,et\,al.\,2023}$ &  [Fe/H]$_{DR3}$ \\

\hline
ASAS J164128-1029.6 & 4337403359591642112 & 12.604 &-0.630$\pm$0.100&-0.810$\pm$0.190& -0.372$\pm$0.258 \\
ASAS J194654+4504.8 & 2079729649857217408 & 12.992 &-0.481$\pm$0.110&-0.690$\pm$0.200 & -0.036$\pm$0.269 \\

\hline
\end{tabular}

\end{table*}

\end{document}